RESEARCH ARTICLE

RUNNING HEAD: Two adaptation processes in a visuomotor rotation task

# Coexistence of two adaptation processes in a visuomotor rotation task

Alexis Berland[1,2,3], Youssouf Ismail Cherifi[2], Alexis Paljic[2], Emmanuel Guigon[3]

[1]Higher Institute of Psychomotor Rehabilitation, ISRP, Paris, France

[2]Mines Paris, PSL Research University, Centre for Robotics, CAOR, Paris, France

[3]Sorbonne Université, CNRS, Institut des Systèmes Intelligents et de Robotique, ISIR, F-75005 Paris, France

Correspondence: Alexis BERLAND (alexisberland1995@gmail.com).

## ABSTRACT

Motor adaptation is a learning process that enables humans to regain proficiency when sensorimotor conditions are sustainably altered. Many studies have documented the properties of motor adaptation, yet the underlying mechanisms of motor adaptation remain imperfectly understood. In this study, we propose a computational analysis of adaptation to a visuomotor rotation task and examine it through an experiment. Our analysis suggests that two distinct processes contribute to produce adaptation: one which straightens trajectories, and another which redirects trajectories. We designed a visuomotor rotation task in a 3D virtual environment where human participants performed a pointing task using a head-mounted display controller represented by a cursor that was visually rotated by an angular deviation relative to its actual position. We observed that: (1) the trajectories were initially curved and misdirected, and became straighter and better directed with learning; (2) the straightening process occurred faster than the redirection process. These findings are consistent with our computational analysis and disclose a new and different perspective on motor adaptation.

## NEW & NOTEWORTHY

This study investigates the visuomotor rotation protocol using a pointing task in order to compare the adaptation of (1) the movement initial angle, (2) the trajectory length. The experiment shows that these two outcomes adapt on different time scales. To account for this observation, we propose a computational analysis based on the control-estimation framework.

# INTRODUCTION

Our ability to control movement is continually challenged by environmental disturbances and noise. Humans demonstrate a remarkable ability in overcoming these challenges through flexible motor control policies. Flexibility is a property of motor control which allows an online compensation for a perturbation. The notion of adaptation refers to a short-term form of learning aimed at minimizing errors caused by changing conditions and enabling individuals to regain their previous performance levels (1). Unlike flexibility, adaptation involves building an internal representation of the characteristics of the perturbation, allowing the movement to account for the nature of the perturbation even before the movement begins. Numerous researchers have investigated the field of motor adaptation (2). Common methods include force field perturbations, prism deviations, gain factors, and visuomotor rotations (VMR) (3, 4, 5, 6). The VMR method introduces a perturbation by incorporating an angular deviation into the visual feedback of a movement, causing a discrepancy between the visual trajectory and the actual hand movement. When describing adaptation, studies commonly focus on either implicit and explicit processes. Implicit adaptation would be characterized by learning curves and aftereffects, whereas explicit adaptation would involve strategy use without aftereffects (7). This distinction, however, is not clear-cut as learning curves can occur even with strategy use (8).

Computational models are mathematical tools which can be used to elucidate the nature of learning processes. State-space models have been proposed to describe the time course of motor adaptation (9, 10) and have raised interesting ideas about the nature of implicit and explicit adaptation processes (11). Yet, this kind of model is phenomenological and addresses only the shape of adaptation curves. A complementary approach could be to assess the contribution of motor control processes (e.g. how movements are produced and corrected online in the presence of perturbations) to adaptation. For instance, in the framework of the control-estimation architecture (12, 13, 14, 15, 16), a movement is produced by the combined operation of a controller and a state estimator (Fig. 1*A*), and motor adaptation could correspond to changes in the control and/or the estimation process (17).

A target-directed movement under VMR generates a typical pattern of dissociation between actual hand displacement and visual feedback (*before effect*; Fig. 1*B*b), compared to the case of veridical visual feedback (*baseline*; Fig. 1*B*a; note that the panels in Fig. 1B are drawings which illustrate the outcome of the computational analysis described in **Computational Analysis**; we have decided not to provide simulations which would require a full description of a model and would prohibitively lengthen the article, but a simulation code is provided in an external repository, see **Data Availability**). According to the control-estimation model, the *before effect* state is due to the fact that the state estimator is tuned to the veridical visual feedback, but receives a rotated visual feedback, which creates a sensory prediction error (see **Computational Analysis**). Typically, VMR adaptation leads to a pattern in which the actual hand moves straight in the proper direction so that the visual feedback terminates on the target (*adapted*; Fig. 1*B*d). A central question is how the motor control system transits from the *before effect* to the *adapted* state. Two processes seem to be involved in the transition: (1) a process which straightens the trajectory; (2) a process which rotates the trajectory. Interestingly, trajectory straightening is observed in the model when the sensory prediction error is zeroed and the state estimator becomes tuned to the perturbation (*estimation adapted*; Fig. 1*B*c; see **Computational Analysis**). Then, the *adapted* state would be obtained by cancelling target error (distance between the visual hand and the target at the end of the movement; Fig. 1*B*, c to d), leading to a rotation of the trajectory. Note that we make no specific claim about how the target error is cancelled during the adaptation (e.g. through goal or movement redirection). In this framework, the restricted exploitation of the target error is not sufficient to account for adaptation since there is no target error in the *before effect* state (Fig. 1Bb), the straightening process being necessary to create a target error. The proposed scenario describes processes, but not how exactly they interact with each other.

The goal of this study is to obtain some experimental support to an adaptation scenario based on the interaction between trajectory straightening and rotation. We conducted a VMR pointing experiment in a Virtual Reality Head-Mounted Display (VR-HMD) environment. Virtual reality devices have the potential to study human motor control by immersing individuals in realistic task environments, promoting an ecological approach. Prior studies have shown that both screen-based VR and VR-HMD can be used to present visuomotor discrepancies (18, 19, 20, 21). We measured the time course of changes in trajectory length as a correlate of the straightening process, and initial trajectory angle as a correlate to the rotation process.

## COMPUTATIONAL ANALYSIS

The control-estimation framework (described here in the linear quadratic case) provides a solution to the following problem: given an object with dynamics

$$\dot{x}(t) = Ax(t) + Bu(t) + n_{dyn}(t)$$

where $x$ is the state vector of the object, $u$ a control vector, $A$ and $B$ matrices, and $n_{dyn}$ noise on the dynamics, and an observation process

$$y(t) = Hx(t) + n_{obs}(t)$$

where $y$ is the observation vector, $H$ the observation matrix, and $n_{obs}$ noise on the observation, find a control law $u(t)$ ($t$ in $[t_0; t_f]$) such that

$$J = \int_{t_0}^{t_f} \|u(t)\|^2 dt$$

is minimum, and $x(t_0) = x^0$, and $x(t_f) = x^G$, where $x^0$ is the initial state, and $x^G$ the final, goal state. The solution is given by the control law

$$u(t) = L(\hat{x}(t), x^G)$$

and the state estimation process

$$\dot{\hat{x}}(t) = A\hat{x}(t) + Bu(t) + K(t)\left(y(t) - \widehat{H}\hat{x}(t)\right)$$

where $\hat{x}$ is the state estimate, $\widehat{H}$ an internal model of the observation matrix, and $K$ the Kalman gain. The quantity

$$y(t) - \widehat{H}\hat{x}(t) = Hx(t) - \widehat{H}\hat{x}(t)$$

is called a sensory prediction error (SPE). The formula for $L$ and $K$ are not necessary for our purpose (see 22 for details).

Consider the simple case where the state $x$ describes the hand position in a 2D plane, and is defined by

$$x = \begin{pmatrix} x_1 \\ x_2 \end{pmatrix}$$

The observation matrix $H$ is defined by

$$H = \begin{bmatrix} H_p \\ H_v \end{bmatrix}$$

where $H_p$ is the proprioceptive observation matrix, and $H_v$ the visual observation matrix. In baseline (veridical) condition, proprioception and vision are aligned and

$$H_p = H_v = \begin{bmatrix} 1 & 0 \\ 0 & 1 \end{bmatrix}$$

In the following, we note $H^0$ the observation matrix corresponding to the veridical condition.

Under VMR, $H_p$ remains unchanged, $H_v$ becomes

$$H_v^\theta = \begin{bmatrix} \cos\theta & -\sin\theta \\ \sin\theta & \cos\theta \end{bmatrix}$$

where $\theta$ is the angle of rotation, and $H$ becomes

$$H^\theta = \begin{bmatrix} H_p \\ H_v^\theta \end{bmatrix}$$

Under the veridical condition, $\hat{H} = H^0$ and the SPE is $H^0(x(t) - \hat{x}(t))$. It is close to zero (not exactly zero because of noise), and $x$ and $\hat{x}$ are aligned. Under VMR (before any adaptation), the state estimator is unaware of the perturbation ($\hat{H} = H^0$), the SPE becomes $H^\theta x(t) - H^0 \hat{x}(t)$. It is nonzero, and $x$ and $\hat{x}$ are not aligned, creating the typical pattern observed when a rotation of visual feedback is introduced (Fig. 1Bb). The very origin of this pattern is the fact that the control law is a function of $\hat{x}$, i.e. the control signal is elaborated based on the estimated state but applied to the actual state.

The fundamental observation underlying this study is the fact that, under VMR, when the state estimator is given knowledge of the perturbation, i.e. $\hat{H} = H^\theta$, the SPE becomes $H^\theta(x(t) - \hat{x}(t))$ and is close to zero under the veridical condition. In fact, the VMR is now the new "veridical" condition which does not create a sensory prediction error. The corresponding behavior is a straight movement to the target with a straight visual feedback rotated by the angle $\theta$ (Fig. 1Bc).

In this framework, true adaptation could be described by a redirection of the goal state $x^G$ in the $-\theta$ direction (Fig. 1Bd).

## MATERIALS AND METHODS

### Participants

Thirty-five healthy young adults from Europe and North Africa provided written informed consent and participated in the study. One participant did not complete the full protocol and was removed from the cohort. Consequently, 34 participants were retained with a mean age of 25 (SD = 4.3), including 15 females and 19 males. According to the Edinburgh Handedness Inventory (23) which assesses manual hand preference from -100 (pure left-handed) to 100 (pure right-handed), 33 participants self-identified as right-handed with a mean score of 77.8 (SD = 26.8), and 1 participant identified as left-handed with a -86.66 score. A differential semantic scale assessed participants' expertise and familiarity with the VR setup across 3 dimensions, each measured by 2 questions with 7 propositions. Each dimension was rated on a scale from 2 to 14. The mean scores for expertise and familiarity were: computers 13.3 (SD = 0.9), video games 10.6 (SD = 3.5), and immersive devices like VR 8.3 (SD = 3.4). The total expertise index, obtained by adding these three scores, indicates a good proficiency in the use of numerical devices (mean = 32.2, SD = 6.3). Self-reported data indicated that all participants were free from neurological impairments. Four participants reported mild visual disorders, such as myopia or astigmatism, which were not always corrected with glasses during the experiment. The experiment was approved by Comité d'Éthique de La Recherche at Sorbonne Université (CER-2021-112). All participants were included in a single cohort and assigned identical tasks.

## Apparatus

Participants were seated on a chair without armrests in front of a table. They were equipped with 6 degrees of freedom and stereoscopic VR-HMD device (Meta® Quest 2, elite strap, inter-pupilar distance set to notch 2 or 3, 90 Hz screen refresh rate) and invited to hold a handheld controller (Meta® Quest 2 Touch controllers) in their dominant hand. A virtual environment developed on Unity® (version 2022.3.4f1) was streamed on the HMD using a computer (MSI®, CPU i913900H, GPU RTX 4070 laptop, Microsoft Windows® 11) through an Air Link Wi-Fi 6 connection (ASUS® RT-AX86S). The controller measured hand position along three axes: left-right (z), down-up (y), backward-forward (x).

## Procedure

The virtual environment was a modified version of the default Meta® software development kit small room environment customized with some furniture assets (see **Data Availability** for pictures). A virtual table adapted to the size, position and color of the physical table was configured so that the participant could place his arm on the physical table and see in the HMD the actions he performed on the virtual table. The starting position for the task was fixed to a rest position chosen by the participant. From the starting position, a target (3D sphere, 2 cm diameter) appeared centered at 15 cm along the x-axis (defined arbitrary as the 0° direction in the horizontal x-z plane) and centered at y = 8 cm above the table. To reach the target, the participant controlled a cursor (3D sphere, 1 cm diameter) which moved in the horizontal x-z plane, centered at y = 8 cm above the table and appeared in place of the handheld controller. Considering the volume size of the target and the cursor, the minimum distance to reach the target was 13.5 cm. No part of the arm, hand or handheld controller were visible during the experiment. Before each trial, the cursor would appear to signify that the starting conditions are met. To force participants to produce neither too slow nor too fast movements, we configured a time limit of 1.5 seconds per trial. The target must be reached within the allotted time to complete the trial. If the time limit of 1.5 seconds was exceeded, the trial ended automatically. The start signal was indicated by a bell ring. When it occurred, the participant could control the cursor to reach the target. If the target was reached in the allotted time, a successful sound occurred, otherwise a failure sound occurred. After each trial, the cursor and the target disappeared, and the participant's hand was guided by a dynamic yellow circle to return to the starting position. The pause time between 2 consecutive trials was randomized from 1 to 2 seconds for each new trial. All participants performed 210 trials split into 3 cycles ordered as follows: 50 with no disturbance (*cycle 1: baseline*), 80 with a 45° clockwise rotation on cursor visual feedback (*cycle 2: adaptation*), 80 with visual feedback and removal of the disturbance (*cycle 3: washout*). Following the experiment, participants completed the French version of the Sense of Agency Scale (F-SoAS) (24, 25) which assesses the sense of being the author of his own actions in virtual environments. The French version of the Simulator Sickness Questionnaire (SSQ-FR) (26, 27) was also addressed to assess the level of discomfort felt during the VR experiment. Finally, participants were queried regarding their emotions, their understanding of the experiment, any observed technical glitches, and their perception of the experiment's objectives.

## Data acquisition, processing, and analysis

The handheld controller's position was sampled at 100 Hz. Data acquisition was restricted to the horizontal x-z plane. For each trial, cartesian velocity was calculated from the first-order derivative of the raw x and z positions. The positions and velocity of each trial were low-pass filtered individually using a fourth-order Butterworth filter with a 10 Hz cutoff. Trajectory length (called *length*) was calculated from the start signal to the cursor's edge collision with the target's edge, based on the filtered positions. The initial trajectory angle (called *angle*) was calculated in the following way ($O$ is the origin, $T$ the target): (1) the position $P$ of a local peak velocity was identified after a displacement between 1 and 2 cm; a first angle $\alpha_1$ was calculated as the difference between the directions $OP$ and $OT$; (2) a linear adjustment (type II regression) was calculated on the trajectory between $O$ and $P$; a $R^2$ was obtained and a second angle $\alpha_2$ was calculated as the difference between the trajectory direction and $OT$; (3) if $|\alpha_1 - \alpha_2| < 10°$ and $R^2 > 0.6$, the identification was deemed consistent and the angle was chosen to be $\alpha_1$; otherwise, the trial was removed from the analysis. The mean proportion of conserved trials across the 3 cycles is 0.94 (SD = 0.06) for both the perturbation and washout groups.

Exponential models were fit to the length and angle data for the perturbation and washout cycles for each of the 34 participants according to

$$y = a - b \times e^{-t/c} \quad \text{(Eq. 1)}$$

where $a$ is the *final steady-state,* $b$ the *magnitude* of change from the initial level, $c$ the *time constant* of change, and $t$ the trial number (28). The quality of fit was described by the coefficient of determination

$$R^2 = 1 - \frac{\sum_i (y_i - \hat{y}_i)^2}{\sum_i (y_i - \bar{y})^2} \quad \text{(Eq. 2)}$$

where $\{y_i\}$ are the data, $\bar{y}$ the mean of the data, and $\{\hat{y}_i\}$ the values predicted by the exponential fit.

The time constants for angle and length for each participant were collected for comparison. Normality tests using Shapiro-Wilk and Q-Q plots showed that the dataset of time constants was not normally distributed. Therefore, we used a non-parametric Wilcoxon signed-rank test with a 5% significance level to compare time constants for length and angle during the adaptation and washout cycles. Statistical analyses were conducted with Jamovi® (v2.3.28) and Conda (Python v3.11.5, NumPy, SciPy, Matplotlib).

## RESULTS

### Single participant

Data of a participant close to an average participant (see Fig. 3B, 3C) are shown in Figure 2. Single baseline and adapted trials exhibited straight trajectories with a bell-shaped velocity profile (Fig. 2*A*, *i* and *iii*), whereas the before and after effect trials exhibited curved trajectories with an irregular velocity profile (Fig. 2*A*, *ii* and *iv*). Quantification is based on trajectory angle (*green*) and trajectory length (*blue*) as illustrated in Figure 2*Aiii* (see **Materials and Methods**). The angle and length curve displayed a steady, yet variable behavior in the baseline cycle (Fig. 2B, *yellow* background). During the adaptation cycle (*white* background), the trajectory angle changed progressively and stabilized at 32.4° with a time constant of 7.7 trials (Fig. 2B, *green*), whereas the trajectory length first spiked and then declined, leveling off at 14.3 cm with a time constant of 0.7 trials (Fig. 2B, *blue*). During the washout cycle, the participant showed an aftereffect with an angle time constant of 8.1 trials and a deadaptation of trajectory length, with a time constant of 0.9 trial (Fig. 2B, *gray* background).

### All participants

One participant failed to adapt to the perturbation and was removed from the analysis. Of the remaining 33 participants: (1) 5 had a time constant of angle adaptation larger than the length of the adaptation cycle and were removed from the analysis of the adaptation cycle (28 participants for the adaptation cycle); (2) 2 had a time constant of angle washout larger than the length of the washout cycle and were removed from the analysis of the washout cycle (31 participants for the washout cycle).

In the adaptation cycle, the mean angle curve (*thin green*) stabilized at an average angle of 37.7° with a time constant of 7.1 trials with a trajectory length (*thin blue*) of 14.6 cm with a time constant of 1.7, close to the baseline trajectory length of 13.8 cm (Fig. 3A). In the washout cycle, the mean angle curve stabilized at an average angle of 5.1° with a time constant of 6.6 trials, with a trajectory length of 14 cm with a time constant of 1.1 (Fig. 3A).

The mean time constant ($c$ from Eq. 1) appeared smaller for trajectory length compared to trajectory angle for the adaptation and washout cycles (Fig. 3A, *green* vs *blue thick* curves). This observation was confirmed visually (Fig. 3B,C) and statistically (adaptation: Wilcoxon test $w(28) = 353$, $p = 1.5 \times 10^{-4}$, d = 0.739; washout: $w(31) = 452$, $p = 6.8 \times 10^{-6}$, d = 0.823). There was a large inter-participant variability both in terms of time constant (Fig. 3$B,C$) and final steady-state ($a$ from Eq. 1 for angle and length; insets in Fig. 3$B,C$), and quality of exponential fit ($R^2$, Eq. 2, for angle and length; insets in Fig. 3$B,C$). There was no significant correlation between $c$ and $R^2$ for angle, length, adaptation, and washout.

Outliers were automatically labeled in the construction of the box plots (Matlab boxplot function; Fig. 3$B,C$, *red* '+'). These "a posteriori" outliers fulfilled our "a priori" inclusion criteria so there is no reason to remove them from our analysis. If they are removed, the results remain qualitatively similar.

## DISCUSSION

The results reported in this study provide a descriptive model of VMR adaptation based on two processes with different dynamics: (1) a fast straightening process which converts the curved trajectory observed when the perturbation is first introduced into a straight trajectory; (2) a slower rotation process which redirects the hand to compensate for the perturbation. We discuss these results in relation to previous studies and previous models of VMR.

Nowadays, VMR protocols require participants to perform fast shooting or slicing movements across the target, which prevents online movement corrections. This kind of protocols tends to restrict VMR adaptation to a sole problem of movement redirection (the word being used here in a neutral sense, irrespective of any implicit or explicit/strategic connotation; see below): the shooting movement creates a misdirected visual feedback and its direction needs to be modified. This restricted view has created some confusion. In order to propose an interpretation of adaptation in terms of internal (forward and inverse) models, it has been proposed that adaptation involves a sensory prediction error (SPE) and a target error (29, 30). In this framework, the adaptation scenario could be as follows. A VMR creates a prediction error, variously defined as:

(1) the difference between the expected and observed outcome (29); (2) the mismatch between the predicted and actual sensory consequences of a movement (31); (3) the mismatch between the intended and sensed location of the effector (32); (4) the difference between where the movement was directed and where the cursor appeared (33). This prediction error would update a forward model (34, 35, 31, 36). According to Lee et al. (36), "the output from the updated forward model can then be used to compute the motor command needed to compensate for the perturbation when a task goal is defined". This scenario would correspond to the implicit component of adaptation, the explicit component being driven by a target error (29, 32). These multiple explanations are not clear. The confusion comes from the use of a prediction error which is not, in the strict sense of the predictive coding theory (37), a genuine SPE (i.e. a difference between a sensory input and a prediction about this sensory input). Our computational analysis (see **Computational Analysis**) shows that: (1) the formalism of forward models is intimately related to a genuine SPE; (2) the cancellation of a SPE related to a VMR leads to a change in the state estimation process, the forward model itself remaining unchanged; (3) the cancellation of a SPE related to a VMR leads to an updating of predictions, a phenomenon which remains behaviorally invisible if not specifically assessed (38); (4) the cancellation of a SPE related to a VMR does not lead by itself to a true adaptation (a target error remains) to the VMR.

In this study, we used the classical VMR protocol in which the goal of the participant is to steer the visual cursor towards a target under rotated visual feedback (39, 40). This protocol is better suited to study visuomotor adaptation than the shooting protocol. It shows that adaptation is not only a matter of change in trajectory angle, but also of changes in trajectory shape (e.g. length). It raises the question of why participants adapt in the absence of actual target errors (the cursor eventually stops in the target at the end of the movement). Our computational analysis suggests that different processes subserve changes in trajectory shape (length) and trajectory angle. Our experimental results are consistent with this view. We observed a faster change in trajectory length (in fact, recovery to the baseline length) compared to the change in trajectory angle. This temporal dissociation generates an adaptation state in which the cursor trajectory is almost straight with an unadapted angle, creating a target error that can drive angle adaptation. We do not claim that the angle and length measures are independent, only that they evolve on a different time scale. Correlation matrices across participants between the time constants of the exponential fits for angle and length do not provide evidence for a dependence between the two processes (see **Data Availability**). It might seem surprising and paradoxical that, during the course of adaptation, the participants produced straight cursor movements which do not reach the target. Yet, one should remember that implicit visuomotor adaptation leads to partial adaptation to the rotation angle, which means that the participants stop adapting even if the trajectory of the visual feedback does not reach the target (41). The situation is still motor striking in clamp protocols (42).

Some previous studies reported measures related to the shape of the trajectory during VMR: curvature (figure 6 in 43; figure 4 in 44; figure 3 in 45; figure 2 in 46; figure 1 in 47; figure 3 in 48; figure 6 in 49), length (figure 1 in 50; figure 4 in 51; figure 2 in 52; figure 3 in 53; figure 5 in 54; figure 3 in 55). None of these studies reported a quantitative comparison between the time course of path length and angle adaptation. The only qualitative comparison is found in Contreras-Vidal and Kerick (52) who reported a rapid improvement in movement duration and movement length, and a gradual improvement in initial directional error.

A classical observation in the classical VMR protocol is that during early trials in the rotated environment, the hand is initially directed toward the target and then deviates in the direction opposite to the rotation in such a way that the cursor tends to reach the target. The actual shape of the hand (and cursor) trajectory is smoothly curved for small rotation angles (12 deg; figure 2 in 56) and less smoothly curved (or even not curved at all) and more irregular for larger rotation angles (45 deg; figure 1 in 50). In both cases, the trajectory is longer (more curved, less straight) for early trials than for late trials. Here, our only contention is that the length of the trajectory decreases during the adaptation process, a phenomenon that we call "straightening" because the trajectory is actually straighter after adaptation than before. We make no claim about how straightening (i.e. trial-by-trial reduction in trajectory length) occurs (e.g. reduction in the size or number of corrective movements). We have not used the term "curvature" to describe our data because, most trajectories were made of multiple straight segments. We have not used the "mathematical" curvature to quantify the adaptation time course because it was not a reliable measure due to the straightness of some segments.

Our results are consistent with the existence of multiple adaptation processes (2). Yet our proposal differs from classical dissociations such as implicit/explicit (7) and slow/fast (9). Our experimental protocol was not designed to control for the involvement of implicit and explicit processes. A posteriori, we observed after-effects in the washout cycle which reveal the contribution of an implicit process. It is tempting to say that the straightening process is implicit and the rotation process is explicit. But, in fact, the two processes create after-effects (in length and angle). In our view, the two processes contribute to implicit adaptation and there probably exists an additional, strategic process, not described in the proposed computational framework, to define the total adaptation. The fast/slow dissociation is not relevant here because the (fast) straightening and (slow) rotation processes do not contribute additively to adaptation.

A closely related proposal is found in Flanagan et al. (57) in the framework of grip force/load force coupling. In Flanagan et al. (57), participants grasped an object with a precision grip and were required to move it along a straight line while being perturbed by a velocity-dependent force field. The perturbation led to an initial increase in hand path length and curvature, but a progressive recovery (straightening of the hand path) was observed within ~70 trials. In contrast, a proper coordination between the grip force and the load force necessary for object handling was established in ~10 trials. In the framework of forward and inverse models, the former adaptation is ascribed to changes in an inverse model and the latter to changes in a forward (predictive) model (58). The interpretation proposed by Flanagan et al. (57) has been called into question by Hadjiosif and Smith (59), but recent findings on adaptation to mirror-reversed visual feedback are consistent with dual forward and inverse model adaptation over differing timescales (60). Note that straightening of the hand path is believed to be related to a change in an inverse model in Flanagan et al. (56) (adaptation to the force field; 3) while we claim here that it is related to the update of the prediction term in a state estimator in VMR adaptation. There is no contradiction between these proposals. In fact, a force field perturbation modifies the dynamics of the controlled object but not the observation process (see **Computational Analysis**) and does not create visual or proprioceptive prediction errors. Optimal feedback control simulations show that adaptation of the state estimator to a force field perturbation tends to straighten the trajectory but only partially compensates for the curvature and lengthening induced by the perturbation.

VR-HMD devices should be used carefully in motor control research because they require adaptation as a human-machine interface, which we tried to address with 50 baseline trials. Besides adaptation issues, technical issues can occur. In our study, 13 participants experienced occasional freezing, 7 reported blurry images at the center or edges, and 3 noticed flickering in their peripheral vision. While these issues were reported as minor by participants, they might have affected motor adaptation performance and the balance of explicit markers. Additionally, some participants occasionally reached peak velocities near the target, suggesting insufficient task boundaries to prevent shooting.

## DATA AVAILABILITY

Data are available at https://zenodo.org/records/15606189 (raw data, processed data, participant information, correlation matrices, setup pictures, tutorial and simulation code for visuomotor rotation adaptation using optimal feedback control).

## GRANTS

Île-de-France Regional Health Agency, Grant Number: 657342.

Foundation for Research in Psychomotricity and Civilizational Diseases – Foundation under the aegis of the Fondation de France, Grant.

## DISCLOSURES

No conflict of interest to declare.

## AUTHOR CONTRIBUTIONS

Conceived and designed research: AB, YC, AP, EG

Performed experiment: AB, YC

Analyzed data: AB, EG

Interpreted results of experiments: EG

Prepared figures: AB

Drafted manuscript: AB

Edited and revised manuscript: YC, AP, EG

Approved final version of manuscript: AP, EG

44. **Schaefer SY, Haaland KY, Sainburg RL.** Dissociation of initial trajectory and final position errors during visuomotor adaptation following unilateral stroke. *Brain Res* 1298: 78-91, 2009.

    https://dx.doi.org/10.1016/j.brainres.2009.08.063

    https://pubmed.ncbi.nlm.nih.gov/19728993

45. **Arce F, Novick I, Vaadia E.** Discordant tasks and motor adjustments affect interactions between adaptations to altered kinematics and dynamics. *Front Hum Neurosci* 3: 65, 2010.

    https://dx.doi.org/10.3389/neuro.09.065.2009

    https://pubmed.ncbi.nlm.nih.gov/20130760

46. **Saijo N, Gomi H.** Multiple motor learning strategies in visuomotor rotation. *PLoS One* 5: e9399, 2010.

    https://dx.doi.org/10.1371/journal.pone.0009399

    https://pubmed.ncbi.nlm.nih.gov/20195373

47. **Isaias IU, Moisello C, Marotta G, Schiavella M, Canesi M, Perfetti B, Cavallari P, Pezzoli G, Ghilardi MF.** Dopaminergic striatal innervation predicts interlimb transfer of a visuomotor skill. *J Neurosci* 31: 14458-14462, 2011.

    https://dx.doi.org/10.1523/JNEUROSCI.3583-11.2011

    https://pubmed.ncbi.nlm.nih.gov/21994362

48. **Mutha PK, Sainburg RL, Haaland KY.** Critical neural substrates for correcting unexpected trajectory errors and learning from them. *Brain* 134: 3647-3661, 2011.

    https://dx.doi.org/10.1093/brain/awr275

    https://pubmed.ncbi.nlm.nih.gov/22075071

## FIGURE LEGENDS

**Figure 1.** A model for visuomotor adaptation. *A*: a control-estimation architecture consists of an object to be controlled, a controller, which translates a goal into motor commands, an observation process and a state estimator. The controller generates the commands needed to achieve a goal based on an estimated state. The estimator computes this estimated state using the commands and the observation (sensory feedback). *B*: drawings of unseen hand displacement (*dashed red*) and corresponding visual feedback (*solid black*) for a movement from an inital position (*small circle*) to a target (*large circle*) under different conditions. (*a*) under a veridical visual feedback (baseline) condition. (*b*) under a 45° clockwise visuomotor rotation, before any adaptation. (*c*) under a 45° clockwise visuomotor rotation, after adaptation of state estimation process. (*d*) under a 45° clockwise visuomotor rotation, after adaptation of state estimation process and redirection of the hand at -45°.

**Figure 2.** Data of a participant. *A*: Trajectories and velocity profiles during a baseline trial (*i*), an early adaptation trial (*ii*), a latter adaptation trial (*iii*) and an early washout trial (*iv*), as indicated in *B*. The small and large circles are the origin and target positions, respectively. Green triangles indicate the identified marker for the calculation of trajectory angle. The method for calculation of trajectory angle (*green*) and trajectory length (*blue*) is shown in *iii*. Calibration: position (0.05 m), time (0.1 s), velocity (0.2 m/s). *B*: Changes in trajectory angle (*thin green*) and trajectory length (*thin blue*) during the baseline (*yellow* background), adaptation (*white* background), and washout (*gray* background) cycles. Exponential fits are indicated by *thick green* and *blue* lines. The *thick, black* line is the angle corresponding to a complete compensation of the visuomotor rotation. The *thin, dotted, blue* is the baseline trajectory length. *Red stars* '*' indicate trials excluded from the analysis (see **Materials and Methods**).

**Figure 3.** Data of all the participants (n = 33 in baseline cycle, n = 28 in adaptation cycle, n = 31 in washout cycle). *A*: Same format as Fig. 2*B*. The *thin green* and *blue* lines indicate the mean curves across participants, and the *vertical* lines, +/- one standard deviation between participants. *B*: Violin and box plots of angle and length adaptation time constants from the adaptation cycle ($c$ from Eq. 1). The box plot indicates the median, the 25th and 75th percentiles, and the extreme values not considered outliers. The outliers are labeled by a *red* '+' marker. The circle corresponds to the mean value. Side inset: log10-transformed adaptation time constant for angle and length. Bottom insets: violin and box plots for the final steady-state of angle and length adaptation ($a$ from Eq. 1), and $R^2$ (Eq. 2) of the exponential fit. The *horizontal green* and *blue* lines indicate the expected magnitude of adaptation. *C*: Same as *B* for the washout cycle.

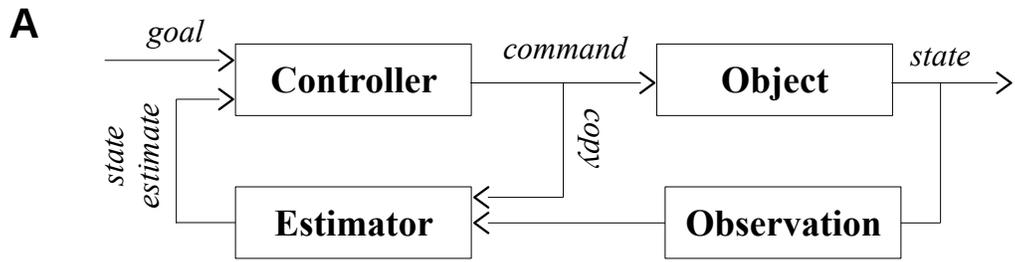
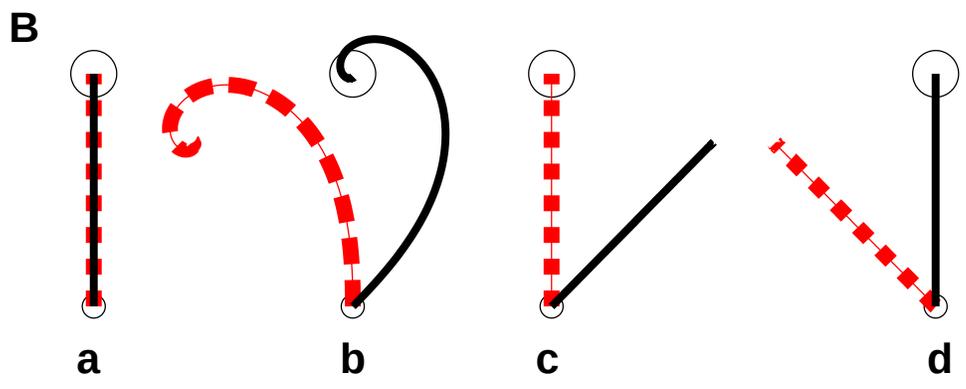

**Figure 1**

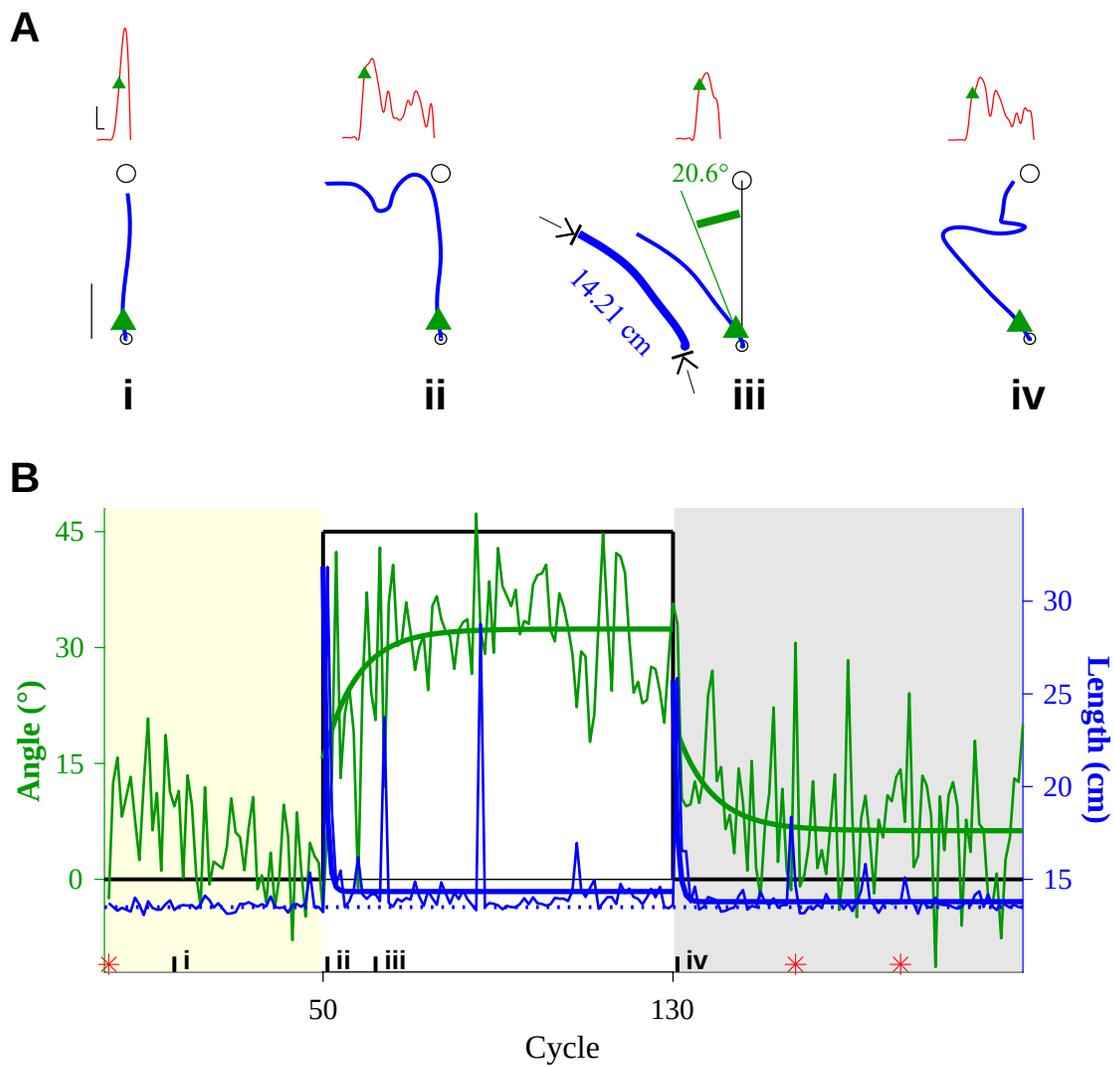

**Figure 2**

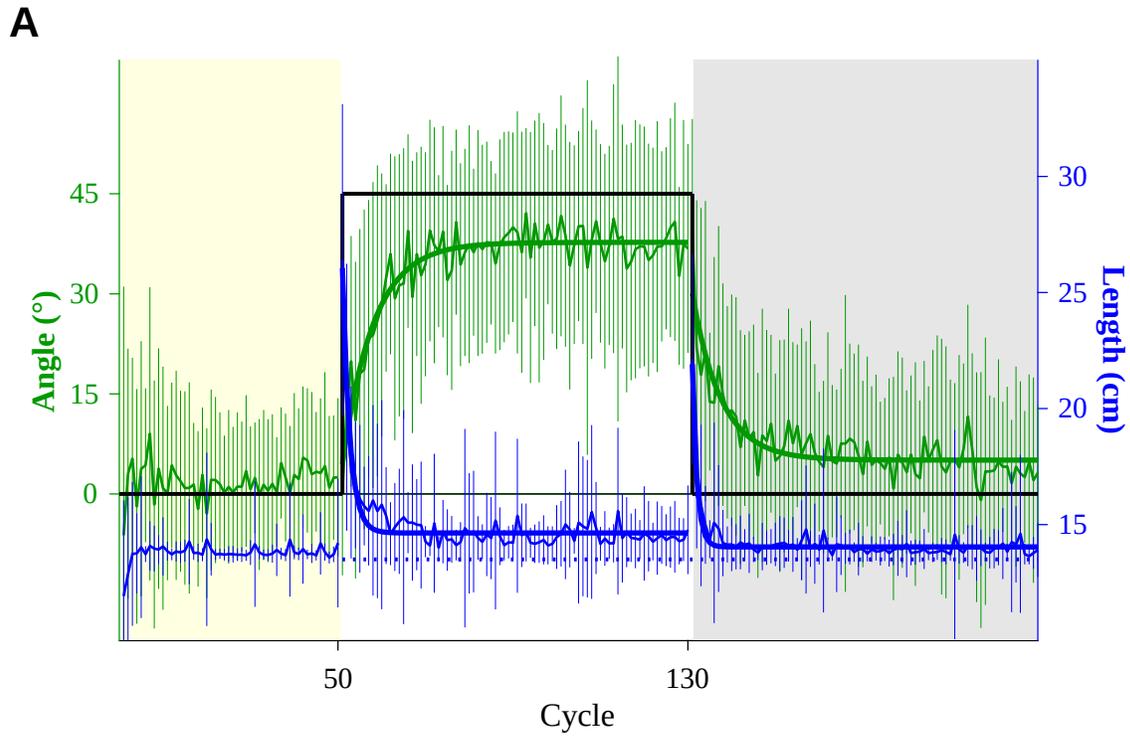
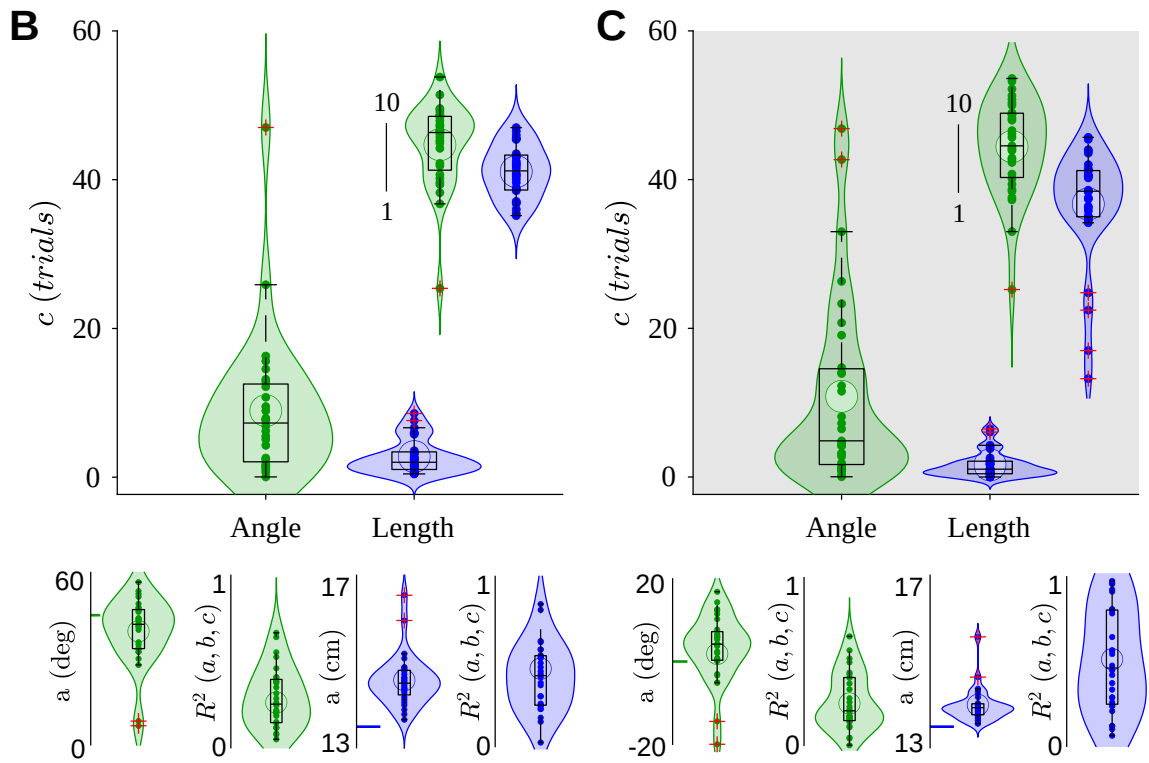

**Figure 3**